\documentclass[12pt]{article}

\catcode`@=12
\begin{document}
\thispagestyle{empty}
\begin{flushright} 
UCRHEP-T327\\ 
March 2002\
\end{flushright}
\vspace{0.5in}
\begin{center}
{\LARGE	\bf New U(1) Gauge Symmetry\\ of Quarks and Leptons\\}
\vspace{1.5in}
{\bf Ernest Ma\\}
\vspace{0.2in}
{Physics Department, University of California, Riverside, 
California 92521, USA\\}
\vspace{1.5in}
\end{center}
\begin{abstract}\
Instead of anchoring the seesaw mechanism with the conventional heavy 
right-handed neutrino singlet, a small Majorana neutrino mass may be 
obtained just as well with the addition of a heavy triplet 
of leptons per family to the minimal standard model of particle interactions. 
The resulting model is shown to have the remarkable property of accommodating 
a new U(1) symmetry which is anomaly-free and may thus be gauged.  There are 
many possible phenomenological consequences of this proposal which may be 
already relevant in explaining one or two recent potential experimental 
discrepancies.
\end{abstract}
\newpage
\baselineskip 24pt

To obtain nonzero neutrino masses so as to explain the observed atmospheric 
\cite{atm} and solar \cite{sol} neutrino oscillations, the minimal standard 
model of particle interactions is often extended to include three neutral 
fermion singlets, often referred to as right-handed singlet neutrinos.  If 
they have large Majorana masses, then the famous seesaw mechanism 
\cite{seesaw} allows 
the observed neutrinos to acquire naturally small Majorana masses. 
On the other hand, there are other equivalent ways \cite{masa,ma98} to realize 
this effective dimension-five operator \cite{wein} for neutrino mass. 
For example, if we replace each neutral fermion singlet by a 
triplet: \cite{ma98,foot}
\begin{equation}
\Sigma = (\Sigma^+, \Sigma^0, \Sigma^-) \sim (1,3,0)
\end{equation}
under $SU(3)_C \times SU(2)_L \times U(1)_Y$, the seesaw mechanism works 
just as well.

It is well-known \cite{mamo} that in the case of one additional right-handed 
singlet neutrino per family of quarks and leptons, it is possible to promote 
$B-L$ (baryon number -- lepton number) from being a global U(1) symmetry to 
an U(1) gauge symmetry.  Here I consider the case of one additional triplet 
of leptons per family, and prove the remarkable fact that a new U(1) symmetry 
exists which is anomaly-free and may thus be gauged.  This discovery leads 
naturally to a number of possible interesting novel experimental consequences.

I assume $SU(3)_C \times SU(2)_L \times U(1)_Y \times U(1)_X$ as a possible 
extension of the standard model, under which each family of quarks and 
leptons transforms as follows:
\begin{eqnarray}
&& (u,d)_L \sim (3,2,1/6;n_1), ~~~ u_R \sim (3,1,2/3;n_2), ~~~ 
d_R \sim (3,1,-1/3;n_3), \nonumber \\ 
&& (\nu,e)_L \sim (1,2,-1/2;n_4), ~~~ e_R \sim (1,1,-1;n_5), ~~~ 
\Sigma_R \sim (1,3,0;n_6).
\end{eqnarray}
There are potentially four Higgs doublets $(\phi_i^+, \phi_i^0)$ with $U(1)_X$ 
charges $n_1-n_3$, $n_2-n_1$, $n_4-n_5$, and $n_6-n_4$.  However, it will turn 
out that three of these four charges are identical, so this model only 
requires the minimum of two distinct Higgs doublets (to be compared with the 
minimum of one Higgs doublet in the standard model).  To allow large Majorana 
masses for $\Sigma$, the Higgs singlet
\begin{equation}
\chi^0 \sim (1,1,0;-2n_6)
\end{equation}
is also added.

To ensure the absence of the axial-vector anomaly \cite{ava}, the following 
conditions are considered \cite{gema}. 
\begin{equation}
[SU(3)]^2 U(1)_X ~:~ 2n_1 - n_2 - n_3 = 0,
\end{equation}
\begin{eqnarray} 
[SU(2)]^2 U(1)_X ~:~ 3 \left( {1 \over 2} \right) n_1 + \left( 
{1 \over 2} \right) n_4  - (2) n_6 = 0,
\end{eqnarray}
\begin{eqnarray}
[U(1)_Y]^2 U(1)_X ~:~ 6 \left( {1 \over 6} \right)^2 n_1 - 3 \left( 
{2 \over 3} \right)^2 n_2 - 3 \left( -{1 \over 3} \right)^2 n_3 
+ 2 \left( -{1 \over 2} \right)^2 n_4 - (-1)^2 n_5 = 0,
\end{eqnarray}
\begin{eqnarray} 
U(1)_Y [U(1)_X]^2 ~:~ 6 \left( {1 \over 6} \right) n_1^2 - 3 \left( 
{2 \over 3} \right) n_2^2 - 3 \left( -{1 \over 3} \right) n_3^2 
+ 2 \left( -{1 \over 2} \right) n_4^2 - (-1) n_5^2 = 0,
\end{eqnarray}
\begin{equation}
[U(1)_X]^3 ~:~ 6 n_1^3 - 3 n_2^3 - 3 n_3^3 + 2 n_4^3 - n_5^3 - 3 n_6^3 = 0.
\end{equation}
Furthermore, the absence of the mixed gravitational-gauge anomaly \cite{mixed} 
requires the sum of $U(1)_X$ charges to vanish, i.e.
\begin{equation}
U(1)_X ~:~ 6n_1 - 3n_2 - 3n_3 + 2n_4 - n_5 - 3 n_6 = 0. 
\end{equation}
Since the number of $SU(2)_L$ doublets remains even (it is in fact unchanged), 
the global SU(2) chiral gauge anomaly \cite{witten} is absent automatically.

Equations (4), (6), and (7) do not involve $n_6$.  Together they allow two 
solutions:
\begin{equation}
({\rm I}) ~ n_4 = -3n_1, ~~~ ({\rm II}) ~ n_2 = {1 \over 4} (7n_1 - 3n_4).
\end{equation}
Using Eq.~(5), solution (I) implies $n_6 = 0$, from which it can easily be 
seen that $U(1)_X$ is proportional to $U(1)_Y$.  In other words, no new 
gauge symmetry has been discovered.

Consider now solution (II).  Using Eqs.~(4) and (6), it implies
\begin{equation}
n_3 = {1 \over 4} (n_1+3n_4), ~~~ n_5 = {1 \over 4} (-9n_1+5n_4).
\end{equation}
Equations (5), (8), and (9) are then \underline {all} satisfied with
\begin{equation}
n_6 = {1 \over 4} (3n_1+n_4).
\end{equation}
This is a remarkable and highly nontrivial result.  In fact, it can be shown 
that the Casimir invariants of the SU(2) representations are such that the 
$only$ solutions to the anomaly-free conditions are with either a singlet, 
i.e. $N_R$, or a triplet, i.e. $\Sigma_R$.

The $U(1)_X$ charges of the possible Higgs doublets are:
\begin{equation}
n_1-n_3 = n_2-n_1 = n_6-n_4 = {3 \over 4} (n_1-n_4), ~~~ n_4-n_5 = {1 \over 4} 
(9n_1-n_4),
\end{equation}
which means that two distinct Higgs doublets are sufficient for all possible 
Dirac fermion masses in this model.  If $n_4 = -3n_1$ is chosen, then again 
$U(1)_X$ will be proportional to $U(1)_Y$.  However, for $n_4 \neq -3n_1$, 
a new class of models is now possible with $U(1)_X$ as a genuinely new gauge 
symmetry.

To summarize, the quarks and leptons transform under $U(1)_X$ as follows:
\begin{eqnarray}
&& (u,d)_L \sim n_1, ~~~ u_R \sim {1 \over 4} (7n_1 - 3n_4), 
~~~ d_R \sim {1 \over 4} (n_1 + 3n_4), \\ 
&& (\nu,e)_L \sim n_4, ~~~ e_R \sim {1 \over 4} (-9n_1 + 5n_4),  
~~~ \Sigma_R \sim {1 \over 4} (3n_1 + n_4).
\end{eqnarray}
The above charge assignments do not correspond to any existing model of quark 
and lepton interactions.  For example, if $n_4=n_1$ is assumed, then
\begin{equation} 
n_1 = n_2 = n_3 = n_4 = -n_5 = n_6,
\end{equation}
which means that $X$ couples 
vectorially to quarks, but its coupling to charged leptons is purely 
axial-vector.  On the other hand, if $n_4 = 9n_1$ is assumed, then
\begin{equation}
n_1 = 1, ~~ n_2 = -5, ~~ n_3 = 7, ~~ n_4 = 9, ~~ n_5 = 9, ~~ n_6 = 3 
\end{equation}
is a solution with $X$ coupling vectorially to charged leptons.

Consider $\nu q$ and $\bar \nu q$ deep inelastic scattering.  It has 
recently been reported \cite{nutev} by the NuTeV Collaboration that their 
measurement of the effective $\sin^2 \theta_W$, i.e. $0.2277 \pm 0.0013 \pm 
0.0009$, is about $3\sigma$ away from the standard-model prediction of 
$0.2227 \pm 0.00037$.  In this model, the $X$ gauge boson also contributes 
with
\begin{eqnarray}
J_X^\mu &=& n_1 \bar u \gamma^\mu \left( {1-\gamma_5 \over 2} \right) u + 
n_1 \bar d \gamma^\mu \left( {1-\gamma_5 \over 2} \right) d \nonumber \\ 
&& +~n_2 \bar u \gamma^\mu \left( {1+\gamma_5 \over 2} \right) u + 
n_3 \bar d \gamma^\mu \left( {1+\gamma_5 \over 2} \right) d + n_4 \bar \nu 
\gamma^\mu \left( {1-\gamma_5 \over 2} \right) \nu.
\end{eqnarray}
Assuming very small $X-Z$ mixing ($|\sin \theta| << 1$), the effective 
neutrino-quark interactions are then given by
\begin{equation}
{\cal H}_{int} = {G_F \over \sqrt 2} \bar \nu \gamma^\mu (1-\gamma_5) \nu 
[\epsilon_L^q \bar q \gamma_\mu (1-\gamma_5) q + \epsilon_R^q \bar q 
\gamma_\mu (1+\gamma_5) q],
\end{equation}
where
\begin{eqnarray}
\epsilon_L^u &=& (1-\xi)\left( {1 \over 2} - {2 \over 3} \sin^2 \theta_W 
\right) + n_1 \zeta, \\ 
\epsilon_L^d &=& (1-\xi)\left( -{1 \over 2} + {1 \over 3} \sin^2 \theta_W 
\right) + n_1 \zeta, \\ 
\epsilon_R^u &=& (1-\xi)\left( -{2 \over 3} \sin^2 \theta_W \right) 
+ n_2 \zeta, \\ 
\epsilon_R^d &=& (1-\xi)\left( {1 \over 3} \sin^2 \theta_W \right) 
+ n_3 \zeta,
\end{eqnarray}
with
\begin{eqnarray}
\xi &=& n_4 \sin \theta \left( 1 - {M_Z^2 \over M_X^2} \right) {g_X \over 
g_Z}, \\ 
\zeta &=& - \sin \theta \left( 1 - {M_Z^2 \over M_X^2} \right) {g_X \over g_Z} 
+ n_4 \left( {M_Z^2 \over M_X^2} \right) {g_X^2 \over g_Z^2}.
\end{eqnarray}

To account for the NuTeV result, i.e.
\begin{eqnarray}
(g_L^{eff})^2 &=& (\epsilon_L^u)^2 + (\epsilon_L^d)^2 = 0.3005 \pm 0.0014, \\ 
(g_R^{eff})^2 &=& (\epsilon_R^u)^2 + (\epsilon_R^d)^2 = 0.0310 \pm 0.0011,
\end{eqnarray}
against the standard-model prediction, i.e.
\begin{equation}
(g_L^{eff})^2_{SM} = 0.3042, ~~~ (g_R^{eff})^2_{SM} = 0.0301,
\end{equation}
consider the following specific model as an illustration:
\begin{equation}
n_1 = 0, ~~ n_2 = -{3 \over 4}, ~~ n_3 = {3 \over 4}, ~~ n_4 = 1, ~~ 
n_5 = {5 \over 4}, ~~ n_6 = {1 \over 4}.
\end{equation}
The central values of the NuTeV measurements are then obtained with
\begin{equation}
\xi = 0.0061, ~~~ \zeta = 0.0038,
\end{equation}
implying that
\begin{equation}
M_X \simeq 10 \left( {g_X \over g_Z} \right) M_Z, ~~~ \sin \theta \simeq 
0.006 \left( {g_Z \over g_X} \right).
\end{equation}
Whereas $M_X \sim 1$ TeV is certainly allowed by the present data, a smaller 
value of $\sin \theta$ is indicated by the precision measurements 
at the $Z$ pole.  A comprehensive numerical analysis of this and the more 
general case of $n_1 \neq 0$ will be given elsewhere \cite{maroy}.

In atomic parity nonconservation, the dominant effect comes from the 
axial-vector coupling of the electron.  In the model defined by Eq.~(29), 
this is given by $(n_4 - n_5)/2 = -1/8$; hence it is rather suppressed. 
Furthermore, the isoscalar vector coupling of the quarks in this model 
also vanishes.  Therefore, the contribution of $X$ is essentially 
negligible and there should be no observable deviation from the prediction 
\cite{apv} of the standard model, in agreement with the most recent data 
\cite{bewi}.

Consider now the anomalous magnetic moment of the muon.  A recent experimental 
result \cite{g-2}, after the latest theoretical corrections \cite{corr}, gives
its deviation from the standard model as
\begin{equation}
\Delta a_\mu = 2.5 \pm 1.6 \times 10^{-9},
\end{equation}
which is only an $1.6\sigma$ effect.  From the standpoint of the proposed 
$U(1)_X$ model, there are two possible contributions.  One comes from the 
$X$ boson which has a vector coupling, i.e. $(n_4 + n_5)/2$, to the muon. 
However, if $M_X \sim 1$ TeV, then this contribution is essentially 
negligible.  The other comes from the extended Higgs sector of this model. 
In particular, the coupling of $(\nu_\mu, \mu)_L$ to $\Sigma_R$ through the 
Higgs doublet with $X$ charge $3(n_1-n_4)/4$ provides two one-loop diagrams: 
one with $\Sigma^-$ and $\bar \phi^0$ as intermediate states, the other 
with $\Sigma^0$ and $\phi^-$.  If all these masses are equal, the former 
contributes with a coefficient of +2 and the latter with a coefficient of 
--1 to $\Delta a_\mu$.  Assuming masses of order 200 GeV, it is 
thus possible to account for Eq.~(32).

It is well-known that given its particle content, the minimal standard model 
does not allow for the unification of gauge couplings.  The addition of 
$\Sigma_R$ in Eq.~(2) does not change the situation.  However, if 
gauge-coupling unification at $M_U \sim 10^{16}$ GeV is desired, one simple 
possibility is to add three charged lepton singlets with only vector 
interactions, i.e. $E_{L,R} \sim (1,1,-1;0)$, as well as an $SU(3)$ octet 
of neutral 
colored fermions, i.e. $\psi_{L,R} \sim (8,1,0;0)$.  It is clear that this 
model would still be anomaly-free, but the evolution equations of the gauge 
couplings would now change, assuming of course that all new particles have 
masses of order $10^2$ GeV.  Generically, the one-loop renormalization-group 
equations for the running of gauge couplings are given by
\begin{equation}
\alpha_i^{-1} (M_1) = \alpha_i^{-1}(M_2) - {b_i \over 2 \pi} 
\ln {M_1 \over M_2},
\end{equation}
where $\alpha_i \equiv g_i^2/4\pi$ and $b_i$ are constants determined by the 
particle content contributing to $\alpha_i$.  Here,
\begin{eqnarray}
&& b_3 = -11 + (3) {4 \over 3} + 4 = -3, \\ 
&& b_2 = -{22 \over 3} + (3) {4 \over 3} + (2) {1 \over 6} + (3) {4 \over 3} 
= 1, \\ 
&& b_Y = (3) {20 \over 9} + (2) {1 \over 6} + (3) {4 \over 3} = 11, \\ 
&& b_X = {1 \over 12} (585 n_1^2 - 282 n_1 n_4 + 177 n_4^2) = (40 ~{\rm if} 
~n_4 = n_1 = 1).
\end{eqnarray}

Using the precision measurements \cite{pdg}
\begin{equation}
\alpha^{-1} (M_Z) = 127.938 \pm 0.027, ~~~ \sin^2 \theta_W (M_Z) = 
0.23117 \pm 0.00016,
\end{equation}
and the relationships
\begin{equation}
\alpha_2^{-1} = \alpha^{-1} \sin^2 \theta_W, ~~~ \alpha_1^{-1} = {3 \over 5} 
\alpha_Y^{-1} = {3 \over 5} \alpha^{-1} \cos^2 \theta_W,
\end{equation}
I find from Eqs.~(33), (35), and (36) that
\begin{equation}
{M_U \over M_Z} = 2.223 \times 10^{14},
\end{equation}
from which $\alpha_3^{-1} (M_Z)$ is predicted by Eqs.~(33) and (34) to be 
8.544, in good agreement with the experimental value \cite{pdg} 
$\alpha_3 (M_Z) = 0.1192 \pm 0.0028$, i.e. $\alpha_3^{-1} = 8.39 
(+0.20/-0.19)$.

In the above, $U(1)_Y$ is normalized as in the standard model, but since the 
normalization of $U(1)_X$ is unknown, $g_X$ cannot be unified in analogy to 
$g_Y$.  This also means that a two-loop analysis of $\alpha_{1,2,3}$ would 
not be possible because it would involve $g_X$.  There is no obvious 
unification symmetry which includes the particle content of this model as 
an anomaly-free subset.

Instead of having one $\Sigma_R$ per family, consider the total of (A) one 
$\Sigma_R$, and (B) two $\Sigma_R$'s for the three families of quarks and 
leptons.  In either case, Eqs.~(4), (6), and (7) are unchanged.  Hence 
solution (II) of Eq.~(10) is still valid, together with Eq.~(11).  The analog 
of Eq.~(5) now implies
\begin{equation}
({\rm A})~~ n_6 = {3 \over 4} (3n_1 + n_4), ~~~ ({\rm B}) ~~n_6 = {3 \over 8} 
(3n_1 + n_4).
\end{equation}
Whereas the analog of Eq.~(9) is still automatically satisfied, that of 
Eq.~(8) is not.  On the other hand, if singlet $N_R$'s are added with $X$ 
charges given as follows:
\begin{eqnarray}
({\rm A}) &:& n_6, ~n_6, -{1 \over 3} n_6, -{5 \over 3} n_6, \\ 
({\rm B}) &:& n_6, {2 \over 3} n_6, -{5 \over 3} n_6,
\end{eqnarray}
the analogs of both Eqs.~(8) and (9) are again satisfied.  Note that in Case 
(A), there are two singlets with $X$ charge $n_6$, and in Case (B), there is 
one such singlet.  This means that the total number of triplets and singlets 
with $X$ charge $n_6$ is always three in each of the three models, thus 
allowing all three neutrinos to acquire small seesaw Majorana masses. 

To conclude, three anomaly-free $U(1)_X$ models have been discovered.  They 
are characterized by having fermions and Higgs bosons beyond those of the 
minimal standard model.  In the simplest case, each family of quarks and 
leptons is supplemented by a triplet of leptons.  In another case, i.e. (A),  
there is only one triplet for the three families, but there are four 
singlets with $X$ charges given by Eq.~(42).  In the third case, i.e. (B), 
there are two triplets and three singlets with $X$ charges given by Eq.~(43). 
If $U(1)_X$ is a relevant gauge symmetry at or near the electroweak breaking 
scale, then it may already be implicated in some recent experimental data 
which show possible deviations from the standard model, such as the NuTeV 
result \cite{nutev} and the muon $g-2$ measurement \cite{g-2}.  Of course, 
the main motivation for studying $U(1)_X$ is not predicated on these potential 
discrepancies, but rather on its fundamental theoretical appeal.  Details of 
other possible phenomenological consequences will be discussed elsewhere 
\cite{maroy}.

This work was supported in part by the U.~S.~Department of Energy under 
Grant No.~DE-FG03-94ER40837.

\newpage
\bibliographystyle{unsrt}

\end{document}